\documentclass[aps,twocolumn,showpacs]{revtex4}
\usepackage{graphicx}
\usepackage{dcolumn}
\usepackage{bm}

\begin{document}
\preprint{}
\title{\bf 
Segregation in aqueous methanol enhanced by cooling and compression
}

\author{L. Dougan$^{(1)}$ } 
\author{R. Hargreaves $^{(1)}$ }
\author{S.P. Bates $^{(1)}$ }
\author{J.L. Finney $^{(2)}$ } 
\author{V. R\'{e}at $^{(3)}$ } 
\author{A.K. Soper $^{(4)}$ } 
\author{J. Crain$^{(1,5)}$ }

\affiliation{$^{(1)}$ School of Physics, The University of Edinburgh,
Mayfield Road, Edinburgh EH9 3JZ, UK}

\affiliation{$^{(2)}$ 
Department of Physics and Astronomy,
University College London, Gower Street, London, WC1E 6BT, UK
}

\affiliation{$^{(3)}$ 
Institut de Pharmacologie et de Biologie Structurale, UMR 5089 - CNRS/UPS, Laboratorie ``RMN et interactions proteines-membranes'', 205 Route de Narbonne, 31077 Toulouse Cedex, FRANCE.
}

\affiliation{$^{(4)}$ 
ISIS Facility, Rutherford Appleton Laboratory,
Chilton, Didcot, Oxon, OX11 OQX, UK
}

\affiliation{$^{(5)}$ IBM TJ Watson Research Center, 1101 Kitchawan Road, Yorktown Heights, New York, 10598, USA}

\date{\today}

\begin{abstract} 

Molecular segregation in methanol-water mixtures is studied across a wide 
concentration range as a function of temperature and pressure. Cluster 
distributions obtained
from both neutron diffraction and molecular dynamics simulations point to 
significantly
enhanced segregation as the mixtures are cooled or compressed. This 
evolution toward greater
molecular heterogenity in the mixture accounts for the observed
changes in the water-water radial distribution function and there are indications also of a change in the topology of the water clusters. The observed behavior is consistent with an approach to an upper critical solution point. Such a point would appear to be ``hidden'' below the 
freezing line, thereby precluding observation of the two-fluid region. 

\end{abstract}


\maketitle

\section{Introduction}
Amphiphiles are a particularly important class of molecule containing both hydrophobic and hydrophilic domains with competing solubility properties. Association of non-polar moities leads to the formation of micelles and other modulated structures. Due to the local density variations inherent to many self-assembled amphiphilic structures, simple equations of state fail to give a complete description of phase equilibria. The simplest amphiphiles 
(the primary alcohols such as methanol, ethanol and propanol) 
are widely studied and have been found to be completely miscible in all proportions and at all state points studied. Recently, however, experimental and computational studies on aqueous methanol have revealed unexpectedly complex behavior at medium lengthscales leading to a substantially revised view of miscibility in this prototype aqueous amphiphile\cite{dixit1,dixit2,dixit3}. In particular, molecular-level segregation has been observed with the alcohol agglomerates exhibiting structural details consistent with those expected for a hydrophobically-driven system\cite{dixit1,dixit3}. Moreover, percolating clusters have been found for both components in a certain concentration range over which many thermodynamic functions and transport properties reach extremal values \cite{dougan}.  

Despite the intense activity and success in studying these model systems at room temperature and pressure, there have been no systematic investigations aimed at mapping out the behavior of these observed extended structures under non-ambient conditions. Even in systems which exhibit clear miscibility gaps, the pressure dependence of the critical solution temperature can increase, decrease or remain constant\cite{schneider} and little information exists on molecular-level structure. 
Such measurements on the model methanol-water mixture are needed to develop and refine molecular-level models of the entropic and enthalpic factors governing the phase behavior of aqueous amphiphiles. These models are potentially of wider significance to areas such as membrane and protein stability. 

Partial miscibility is a common feature of binary liquid phase equilibria in which a mixture separates into two phases of different compositions\cite{schneider} depending on temperature and pressure. This behavior follows directly from the {\em Gibbs phase rule} and is 
contained within simple molecular thermodynamic models such as Bragg-Williams theory. Typically, an immiscible region terminates at an {\em upper critical solution temperature} (UCST), above which the mixture is fully miscible. In some hydrogen-bonded systems, however, further cooling leads to re-entrant miscibility and a closed-loop gap in the phase diagram appears\cite{jackson,davies,marsh}.
We therefore report here an attempt to explore structural properties of methanol-water mixtures far from the ambient state point. We use a combination of neutron diffraction with comprehensive isotope substitution and classical molecular dynamics simulations. The specific objective of the work is to identify the separate effects of temperature and pressure on the structures formed in these solutions, the combined effects of temperature and pressure  and to comment on the 
nature of intermolecular contacts in these solutions.

\section{Experimental and simulation details}

Protiated and deuteriated samples of methanol and water were obtained
from Sigma-Aldrich and used without additional purification. Neutron
diffraction measurements were performed on the SANDALS time-of-flight
diffractometer at the ISIS pulsed neutron facility at the Rutherford
Appleton Laboratory in the UK. The liquid samples were contained in
flat plate cells constructed from a Ti-Zr alloy from which coherent
scattering is negligible.  These were mounted on a closed cycle
refrigerator, and neutron diffraction measurements were made at a
number of different temperatures and pressures(Table
\ref{tab:epsr_details}). For the high pressure experiments, the sample
was contained in several 1.5 mm cylindrical channels cut into a flat
TiZr plate.  Pressure was applied using an intensifier. The high pressure experimental arrangement has been described in detail in a previous publication\cite{postorino}.  The data were
corrected for attenuation, inelastic and multiple scattering using the
ATLAS programe suite\cite{aks1}. The differential scattering
cross-section for each sample was obtained by normalising to a
vanadium standard sample. A total of 7 isotopically distinct samples were measured for methanol mole fractions 
$x=0.27$, $x=0.54$ and $x=0.70$.  These were respectively (i) $CD_3OD$
in $D_2O$; (ii) $CD_3OH$ in $H_2O$; (iii) a 50:50 mixture of (i) and
(ii); (iv) $CH_3OD$ in $D_2O$; (v) a 50:50 mixture of (i) and (iv);
(vi) $CH_3OH$ in $H_2O$; and (vii) a 50:50 mixture of (i) and (vi).
For $x=0.05$, 5 samples were measured, namely (i),(ii),(iii),(vi)
and (vii) and for $x=0.50$ (i),(ii),(iii),(iv)
and (v). These procedures lead to a structure factor $F(Q)$ having
the form $F(S_{\rm HH}(Q),S_{\rm XH}(Q),S_{\rm XX}(Q))$ where
S$_{HH}$(Q) relates to correlations between labelled atoms and S$_{XH}$(Q)
and S$_{XX}$(Q) are the two composite partial structure factors which
give the remaining correlations between other types of atoms (X) and
the labelled atom type (H) in the form of a weighted sum of individual
site-site partial structure factors.

\begin{figure}
\centering
\includegraphics[angle=270,width=10cm]{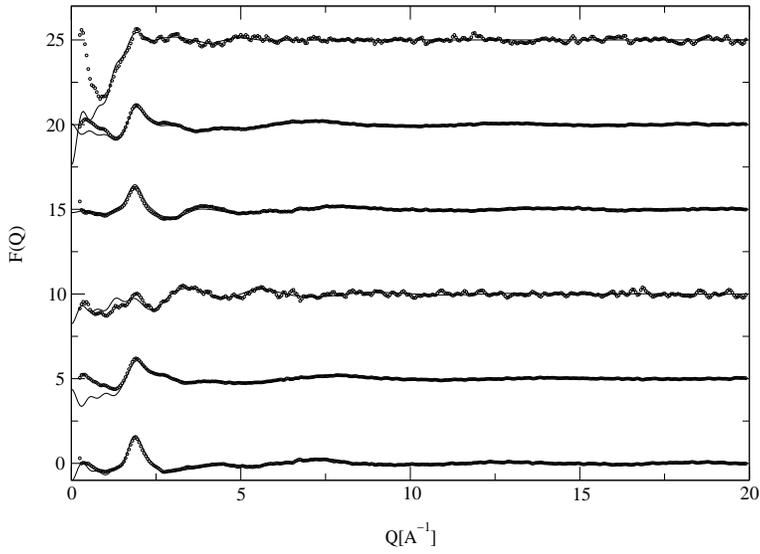}
\caption{Typical example of the fits (lines) obtained by the 
EPSR analysis compared to the original data
(circles). The data shown in this case ($x = 0.50$ at 200K and 2kbar) are the interference
differential scattering cross-sections for the samples (i) through
(vi) described under Methods. Discrepencies are observed in the low
$Q$ region. These are caused by difficulties in removing completely the effect
of nuclear recoil from the measured data. However this recoil effect
is expected to have only a monotonic dependence on $Q$ and so is
unlikely to influence the model structure to any significant extent.}
\label{fig:structurehigh}
\end{figure}

Diffraction data is analyzed using the Empirical Potential Structure
Refinement procedure(EPSR)\cite{sop1}. According to this method, a
three-dimensional computer model of the solution is constructed and
equilibrated using interaction potentials taken from the
literature. The charges and Lennard-Jones constants from the SPC/E potential of Berendsen et al \cite{spce} were used for the water molecules. The H1 potential of Haughney et al \cite{methpot} was used for the methanol molecules. Methanol-water interactions were simulated by Lorentz-Berthelot mixing rules\cite{Allen}. Information from the diffraction data
is then introduced as a constraint whereby the difference between
observed and calculated partial structure factors enters as a
potential of mean force to drive the computer model (via Monte Carlo
updates of atomic positions) toward agreement with the measured data.
This procedure results in an ensemble of three-dimensional molecular
configurations of the mixture exhibiting average structural
correlations that are consistent with the available diffraction data.
A total of 600 molecules (methanol and water) are contained in a cubic
box of the appropriate dimension to give the measured density of each
solution at the appropriate temperature(Table \ref{tab:epsr_details}). Periodic
boundary conditions are imposed. A comparison between the experimentally-measured partial structure
factors and those generated from the ensemble-averaged EPSR with 10000
configurations is shown in Fig \ref{fig:structurehigh}.

\begin{table}
\begin{center}
\begin{tabular}{|c|c|c|c|c|c|c|c|} \hline

Mole fraction & Temp. & Pressure & Total No. & No. of methanol & No. of water & No. density & Box Size \\
$x$ & /$K$ & kbar & molecules & molecules & molecules & /  atoms/$\AA^{3}$ & /\AA \\
\hline
0.27 & 293 & amb & 600 & 162 & 438 & 0.0967 & 28.69  \\ \hline
0.27 & 238 & amb & 600 & 162 & 438 & 0.0967 & 28.69       \\ \hline
0.50 & 200 & amb & 600 & 300 & 300 & 0.1026 & 29.74  \\ \hline
0.50 & 200 & 2.0 & 600 & 300  & 300  & 0.1158 &   28.57  \\ \hline
0.54 & 298 & amb & 600 & 324 & 276 & 0.0955 & 30.26 \\ \hline
0.54 & 260 & amb & 600 & 324 & 276 & 0.0975 & 30.52 \\ \hline
0.54 & 190 & amb & 600 & 324 & 276 & 0.1000 & 30.73 \\ \hline
0.70 & 293 & amb & 600 & 420 & 180 & 0.0930 & 32.04 \\ \hline

\end{tabular}
\end{center}

\caption{Parameters of the methanol-water mixtures used in the Empirical Potential Structural Refinement.}
\label{tab:epsr_details}

\end{table}

\begin{table}
\begin{center}

\begin{tabular}{|c|c|c|c|c|c|c|c|} \hline

Mole fraction & Temp. & Pressure & Total No. & No. of methanol & No. of water & No. density & Box Size \\
$x$ & /$K$ & kbar &molecules & molecules & molecules & /  atoms/ $\AA^{3}$ & /\AA \\
\hline

0.27 & 298 & amb & 600 & 162 & 438 & 0.0967 & 28.69 \\ \hline
0.27 & 298 & 2.0 & 600 & 162 & 438 & 0.1142  & 27.68 \\ \hline
0.50 & 200 & amb & 600 & 300 & 300 & 0.1027& 29.74 \\ \hline
0.50 & 200 & 2.0 & 600 & 300 & 300 & 0.1158& 28.57 \\ \hline
0.50 & 298 & 2.0 & 600 & 300 & 300 & 0.1080& 29.24 \\ \hline
0.54 & 298 & amb & 600 & 324 & 276 & 0.0955& 30.73 \\ \hline
0.54 & 190 & amb & 600 & 324 & 276 & 0.1000& 30.26 \\ \hline
0.70 & 298 & amb & 424 & 297 & 127 & 0.0934& 28.50 \\ \hline
0.70 & 298 & 2.0 & 424 & 297 & 127 & 0.107  & 27.22 \\ \hline

\end{tabular}
\end{center}

\caption{Parameters of the methanol-water mixtures used in the Molecular Dynamcics simulations using {\tt DL\_POLY}.}
\label{tab:md_details}

\end{table}

We have performed a series of classical molecular dynamics simulations using
the {\tt DL\_POLY} code \cite{dlpoly} employing previously tested
intermolecular potentials for both methanol \cite{pere0101} and water
\cite{levi9701}. Both molecular species are modelled using a
fully-flexible, all-atom approach with specific van der Waals terms for
each atom type.  Our previous studies \cite{dougan}, \cite{bate0401} have shown
this code and these potentials can predict the local and extended
structural and dynamical behaviour of these mixtures across the
composition range in close agreement with empirical observation. The
simulations were run to produce 2ns trajectories using 0.5fs timestep
with an equilibration time of over 0.5ns. The sampling interval on the 
trajectories were every 0.1ps. Due to this wealth of data, most of the 
subsequent statistical analysis was done only on the second half of the 
trajectory giving around 10000 data points. Details of systems used in simulations are given in Table \ref{tab:md_details}.

Temporal averages of the molecular dynamics simulations and
configurational averages of the EPSR-generated ensembles were then
interrogated to extract complete sets of intermolecular structural
correlations including information on short-range (solvation) and
medium-range (clustering) structure. In the case of water, clusters
are defined by those molecules that participate in a continuous
hydrogen-bonded network. Two water molecules are considered to be
hydrogen-bonded if the inter-oxygen contact distance is less than
approximately 3.5 \AA, the radial distance of the 
first minimum of the $g_{Ow-Ow}(r)$ pair
correlation function for both EPSR and MD ensembles. The definition
of a cluster is ambiguous in the case of methanol and can be made two different ways. The first is on the basis of hydrogen-bond connectivity, i.e. if constituent methanol oxygen atoms are less the distance of the first minimum in the $g_{O-O}(r)$ pair correlation function.  The second is through methyl group
association, where two methanol molecules are assigned
to the same cluster if the C-C distance is less than the minimum following the first peak determined from the $g_{C-C}(r)$ pair correlation function (which is approximately
5.7 \AA). The former is more common in pure methanol whereas the latter
connectivity type is 
believed to be characteristic of molecular association that is driven by the 
hydrophobic interaction. These two types of clusters are subsequently referred
to as polar contact clusters and non-polar contact clusters.

\section{Results}

\subsection{Cluster distributions at low temperature}

\begin{figure}
\includegraphics[width=15cm]{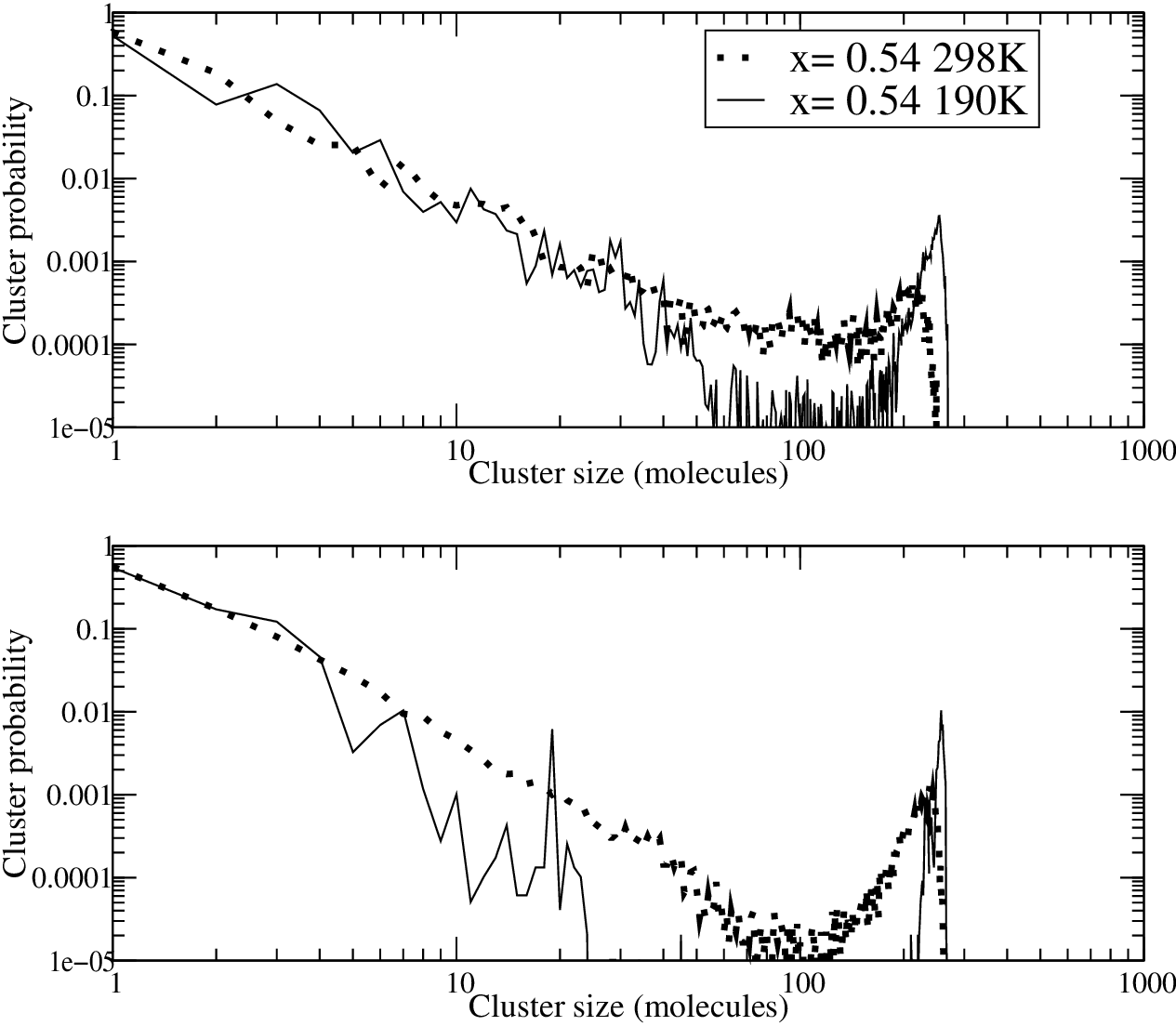}

\vspace{6cm}
\caption{The effect of cooling on water cluster distributions for x=0.54 from EPSR analysis (top) and MD simulations (below)}
\label{fig:5446clusterboth}
\end{figure}

We have already demonstrated in a previous publication\cite{dougan}
that the methanol-water system exhibits significant micro-segregation
across a wide range of compositions, forming localised pockets of 
a single species of varying size and topology. These clusters
are characterised according to the criteria outlined previously. 
The behavior of water cluster distributions on cooling
at mole fraction $x=0.54$ are shown in Fig \ref{fig:5446clusterboth}. The number of clusters containing $i$ molecules is plotted as a fraction of total number of clusters, $M(i)/M$ (where $M= \sum_{i}M(i)$) against the cluster size $i$.
The cluster distributions show an enhanced probability of the largest clusters
on cooling, at the expense of medium-sized (100 molecules or so) clusters. 
There are slight 
differences in the experimental and simulation-derived
distributions plots; 
the depleted region of medium-sized clusters in the MD simulations is wider
and the enhanced peak of the largest clusters is narrower than the corresponding
features in the experimental plot. However, the plots do show the same
trends, with the main features being that the system 
exhibits larger water clusters and these clusters are more frequently 
present upon cooling. This is consistent with increased segregation of the 
two components upon cooling. The same trends are seen in the EPSR analysis of a  mole fraction $x=0.27$ mixture, although the effect is less marked since the water clusters are already bigger at this concentration. Results of MD simulations at other compositions (not shown) show the same 
behaviour in water cluster size distribution on cooling. 

\subsection{Cluster distributions at elevated pressure}

\begin{figure}
\centering
\includegraphics[width=12cm]{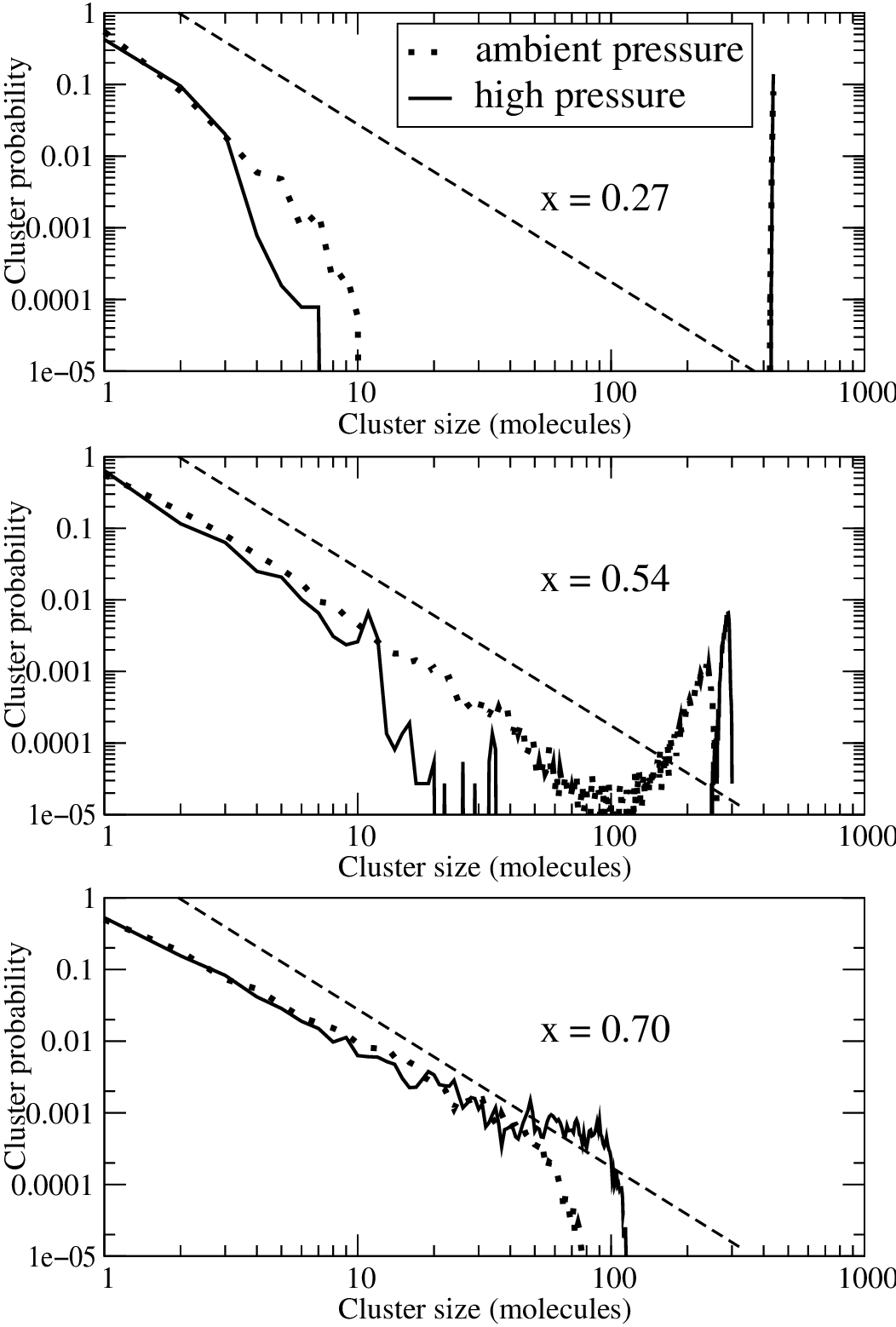}

\caption{The effect of compression on MD simulation cluster distributions for  mole fraction x = 0.27 (top), x = 0.54 (middle) and x = 0.70 (bottom) for water clusters. Ambient conditions are shown as dotted lines and high pressure shown as solid lines. The  predicted power law $n_s \approx s^{- 2.2}$ for  
random percolation on a 3-d cubic lattice\cite{jan} is shown as a dashed line.}
\label{fig:clusterpressMD}
\end{figure}

The effect of compression to 2kbar on the ambient temperature
methanol-water cluster distributions determined by analysis of MD simulations
is shown in Fig.
\ref{fig:clusterpressMD} at several concentrations, 
$x=0.70$, $x=0.54$ and $x=0.27$. Also shown is the  predicted power law $n_s \approx s^{- 2.2}$ for  
random percolation on a 3-d cubic lattice\cite{jan}. 
The results for water clusters in
the system and the effect of increased pressure is most evident on the solutions of 
methanol mole fraction $x=0.7$ and $x=0.54$. 
In both cases, the size and probability of occurrence of the largest water clusters
is increased. 
  Particularly striking is the
case of the most concentrated ($x=0.7$) mixture where compression of 2kbar
changes the water connectivity from isolated non-spanning clusters to a percolating network.
The effect on the most dilute solution ($x=0.27$) is less obvious as the ambient pressure 
results indicate that this solution already comprises very large water clusters, in excess
of the theoretical limit for random percolation. 
We thus find that the effect of pressure is
also to enhance segregation at all concentrations studied. The qualitative effect
of compression is therefore very surprising, being the
same as the effect of cooling. This result is unexpected and contradictory to the general expectation that compression should have had the opposite, destructuring effect.
The methanol cluster distributions, both non-polar and 
polar (as defined previously) were also explored. In the case of the non-polar 
clusters, it is difficult to discern a notable effect of compression: the cluster distributions 
of even the ambient pressure systems 
are dominated by large clusters, often comprising all the methanol in the system. Likewise for the 
polar clusters, of which there are only a relatively small number
of small clusters, the effect of compression is rather small.

\subsection{Cluster distributions at elevated pressure and reduced temperature}

\begin{figure}
\centering
\includegraphics[width=15cm]{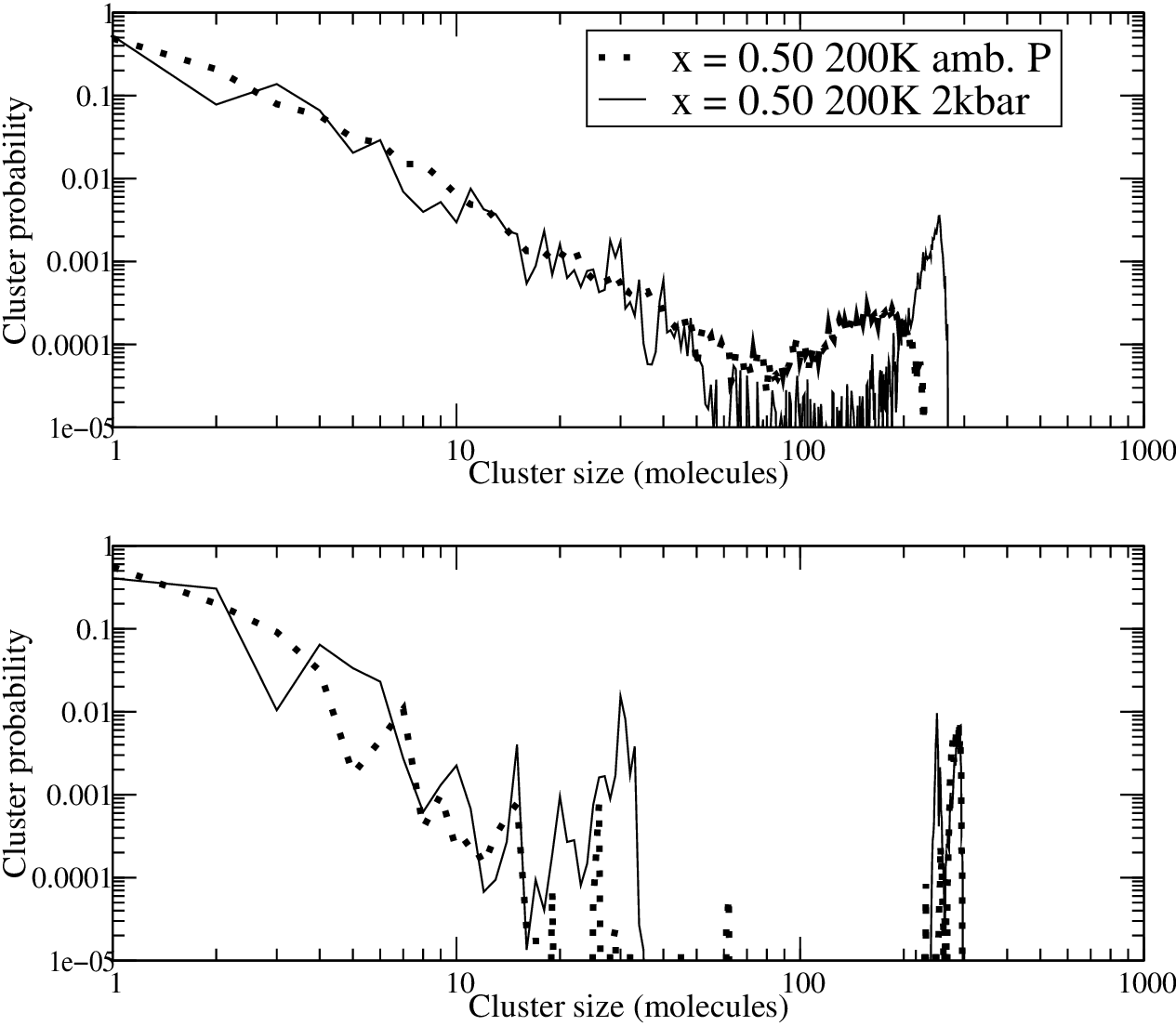}
\vspace{4cm}

\caption{The effect of compressing a cooled system for mole fraction $x=0.50$ on water cluster distributions from EPSR analysis (top) and MD simulations (below)}
\label{fig:PTclusterboth}
\end{figure}

We also consider the effect of compressing the cooled system. 
Figure \ref{fig:PTclusterboth}  shows the corresponding results for water clusters from EPSR
analysis of the neutron data and MD simulations obtained at $x=0.50$ and
T=200K at ambient pressure and 2kbar.
The effect of compressing a cooled system appears to be a further enhanced probability of 
larger water clusers, as was seen previously for the effects of lowered temperature or compression
alone. The effect is clearest from the EPSR analysis, where once again the depletion of medium
size water clusters is evident. The cluster distribution shows an enhanced probability of the largest clusters
on compression, at the expense of these medium-sized clusters. The data from the MD simulations are broadly consistent with this picture,
and appear to show a bimodal distribution of large cluster sizes, centered around 250 and 280 
water molcules. At this composition, there are only 300 water molecules in the system, indicating
that these two peaks actually pertain to the same cluster, which absorbs or sheds smaller
clusters during the course of the simulation. The prominent peak around a cluster size of 30 is a strong
candidate for involvement in this process. 
We return to the differences between EPSR and MD data in the discussion. 

\subsection{Local structure}

\begin{figure}
\centering
\includegraphics[angle=270,width=15cm]{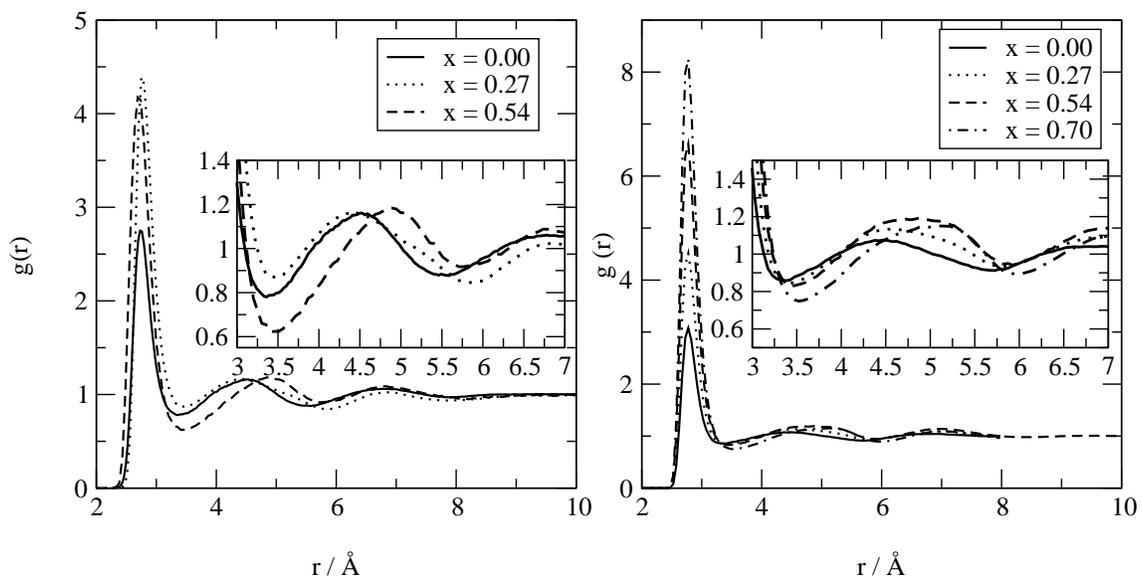}
\caption{Water radial distribution funcations $g_{Ow-Ow}(r)$ from both EPSR analysis (left) and MD simulations (right) at different concentrations. Inset shows detail of second peak.}
\label{fig:OwOwconcB}
\end{figure}

In this section we examine the local structure focussing on the pressure and
temperature behavior of the water oxygen  radial distribution function $g_{Ow-Ow}(r)$.
We first consider the 
effects of concentration on this RDF at ambient temperature and pressure. 
Fig \ref{fig:OwOwconcB} shows data
 from both EPSR and MD analyses. Even at the lowest concentration, 
with mole fraction
$x=0.27$ 
we find that a perturbation
to $g_{Ow-Ow}(r)$ compared to that of pure water 
is 
evident. The location of the second maximum shifts $\approx 0.2 \AA$ 
to higher $r$ from the MD simulations. 
At increasing concentrations we find, for both experiment and MD simulation, 
progressively larger shifts in the 2nd peak postion to larger r values. This shift to higher r implies that the water clusters are becoming less like bulk water, which is supported by cluster distribution\cite{dougan}, which show smaller average water cluster sizes with increasing methanol concentration. We assume this is a consequence of the water
being confined to increasingly smaller domains by the surrounding methanol hydroxyl groups
and consequent interfacial tension.  

\medskip

\begin{figure}
\centering
\includegraphics[angle=270,width=15cm]{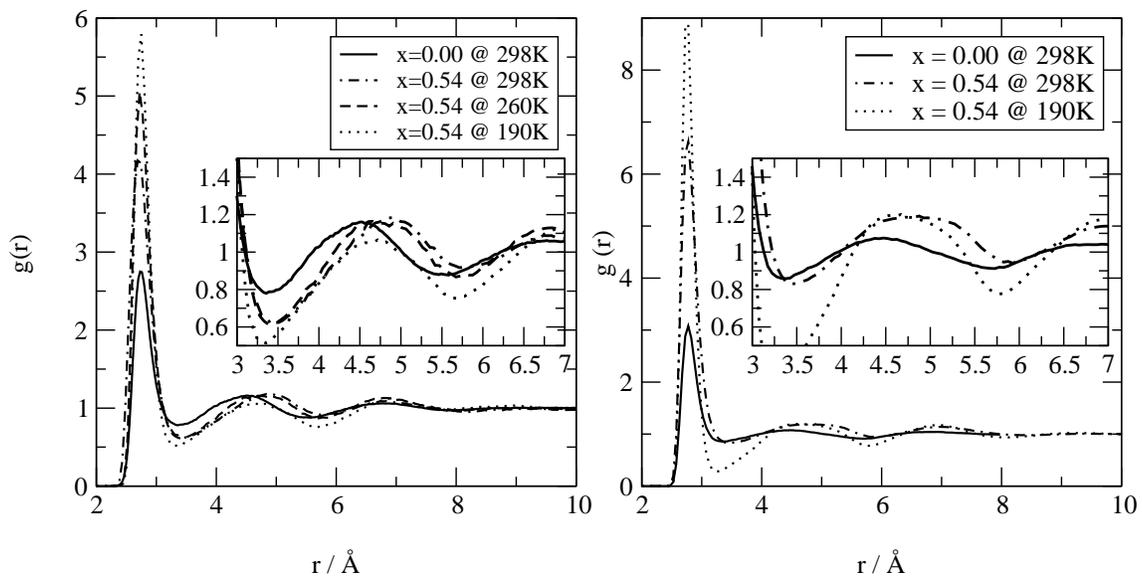}
\caption{The effect of cooling on the water radial distribution function $g_{Ow-Ow}(r)$ from both  EPSR analysis (left; x=0.54 at 298K, 260K and 190K) and MD simulation (right; x=0.54 at 298K and 190K).}
\label{fig:OwOwtemp5446}
\end{figure}

The effect of cooling on $g_{Ow-Ow}(r)$ is also shown in Figure \ref{fig:OwOwtemp5446}.
At $x=0.27$ (not shown) we find that the main features of the distribution are sharpened perhaps indicative
of reduced dynamic disorder but that
there are no other significant changes. By contrast, at the higher alcohol concentration of $x=0.54$,
comparatively large structural
perturbations induced by the presence of the alcohol molecules are partially reversed on cooling
and the displaced second shell peak in the radial distribution function
moves back towards its original position (i.e. that of pure water). 
As the position of this 2nd peak is generally associated with tetrahedrality of the local water structure, both the EPSR analysis of the experimental data and the MD simulations suggest that the previously perturbed tetrahedral structure of the water is recovered on cooling.

\begin{figure}
\centering
\includegraphics[angle=270,width=15cm]{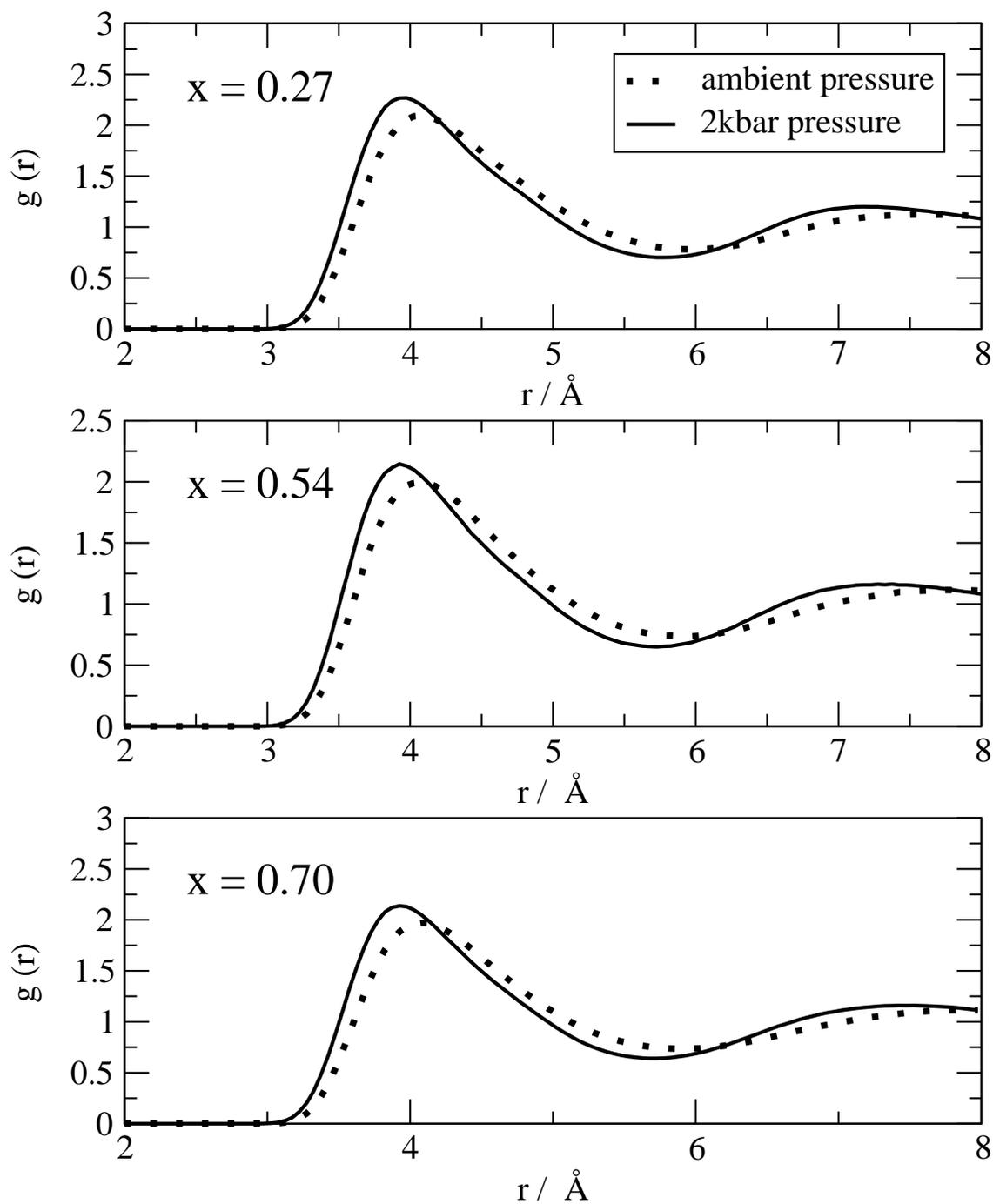}
\caption{Methanol $g_{C-C}(r)$ radial distribution functions from MD simulations (mole fraction x=0.27; 0.54 and 0.70).  }
\label{fig:CCpRdfMD}
\end{figure}

We next consider the effect of compression on the local structure
at ambient temperature. MD simulations of a $x=0.54$ mole fraction solution
do not show any obvious change to the 
position of the second peak in the $g_{Ow-Ow}(r)$ (in contrast to the data in Fig
\ref{fig:OwOwtemp5446} for low temperature). 
Interestingly, it is
the methanol $g_{C-C}(r)$ radial distribution function which is perturbed most
significantly, and this is shown in Fig \ref{fig:CCpRdfMD}.
At three different concentrations $x$=0.70, $x$=0.54 and $x$=0.27, the first
and second peaks in $g_{C-C}(r)$ are seen to shift to lower $r$ values. This
shift is approximately the same for all concentrations as the methanol content is increased. This indicates
that the methyl groups are squeezed together as the pressure is
increased. It appears that it is the methanol which is most responsive to the
pressure.
  
\begin{figure}
\centering
\includegraphics[angle=270,width=15cm]{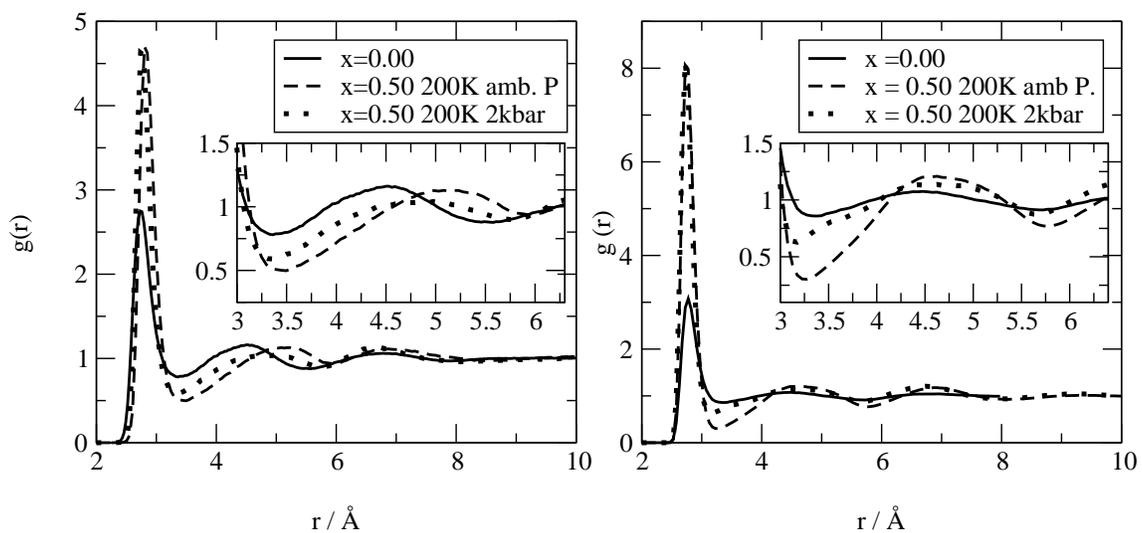}
\caption{Water radial distribution functions $g_{Ow-Ow}(r)$ for mole fraction x=0.50 at 200K, variable pressure from EPSR analysis (left) and MD simulations (right)}
\label{fig:OwOw5050}
\end{figure}

\medskip
Finally we consider the combined effect of reduced temperature and elevated pressure on the local 
structure. 
We might have expected 
the combined effect to be similar to the sum of the 
effects of the consituent parts. However, from both EPSR analysis and MD simulations we find 
(see Fig. \ref{fig:OwOw5050}) that compression of a cooled solution results in 
a further shift to lower $r$ of the second 
 peak in the $g_{Ow-Ow}(r)$. Thus the water appears to partially recover its unperturbed (ie ambient 
pressure and temperature) structure. 
In addition, the first and second peaks in the 
methanol $g_{C-C}(r)$ are shifted to lower $r$, analagous with the situation for the
compression of the solution at ambient temperature.

\section{Discussion}

The prediction from both the EPSR analysis and MD simulations is one 
of enhanced segregation of methanol and water on cooling, evidenced
by the more frequent existence of larger water clusters,
in comparison to room temperature data. 
This 
is the kind of behaviour we would expect on a microscopic scale if the system were moving towards
a phase boundary, characterised by an upper critical solution temperature.
Such immiscibility has not been observed in the methanol-water system, this is because the intervening solid phase precludes access to any possible two-fluid region in methanol-water mixtures.
The clustering behaviour at low temperature provides a consistent framework within which to interpret the
observed variations in local structure, particularly $g_{Ow-Ow}(r)$.
As the temperature is lowered the formation of larger clusters  leads to increased connectivity
of the water domains. Within these growing water clusters the local structure evolves
toward that of bulk water. 
The effect is most obvious with the methanol-rich solutions studied, of mole
fraction $x=0.54$. 

\medskip

Compression of the solution leads to the same effect on the medium-range order of the
system, that is to enhance segregation by formation of larger water clusters. 
However, the local structure shows little if any change in the positions of the 
peaks in $g_{Ow-Ow}(r)$. In contrast, the corresponding RDF for methanol carbon atoms
in systems at elevated pressure 
is displaced to lower $r$ at all concentrations. 
It would seem therefore that the topology of the larger clusters formed by enhanced 
pressure is different to those formed by lowered temperature, since the water contained
within them does not show a significant trend in the RDF back towards that of bulk water 
(as was the case for lowered temperature).

Both the MD simulations and the EPSR analysis predict qualitatively the same trends of 
enhanced clustering as a function of lowered temperature and increased pressure. There
are however differences in the predicted cluster distributions. One possible reason for this
may be the sampling of different average configurations. The EPSR analysis samples Monte Carlo
configurations, whereas the MD samples temporal snapshots. Within the cluster analysis of 
MD simulations, a peak in the cluster distribution plots may occur from either a large 
number of clusters of size $i$ occuring or a relatively smaller number of the same size that persist 
for a long period of time during the simulation. 
It is not surprising therefore that the precise details of the cluster distributions are 
different since
they represent different methodologies of arriving at the same prediction. The qualitative 
trends are clear from both analyses; that clustering, and hence segregation of the two species,
is enhanced by elevated pressure or reduced temperature. A clear example of the differences in distributions is shown in Figure \ref{fig:PTclusterboth}, for the compression of a low temperature $x$=0.54 mole fraction 
solution. The EPSR analysis clearly shows that the size of the largest 
water cluster increases and clusters of that size are more frequently 
found compared to the ambient pressure case. The MD results on the other 
hand are indicative of larger clusters in the sense that they predict 
the existence of two very large clusters (though not simulataneously) 
centered around 250 and 280 molecules. The combined probabilty of these 
clusters is greater than that of the largest clusters present in the 
ambient pressure data.

We note that our results concerning the enhanced segregation at elevated pressure are 
in contrast to the results of Hummer {\it et al} \cite{Hummer} who have concluded that
pressure destabilises the contact configuration of non-polar molecular groups, relative 
to a solvent-separated configuration. These authors then assert that pressure denaturation of proteins 
proceeds by a similar mechanism, that is solvent penetration into a hydrophobic core. In contrast to this our results from diffraction measurements and simulations indicate the hydrophobic groups get pushed closer together with pressure. This difference may be due to the consequence of having an amphiphile in solution rather than a simple hydrophobe.

%

\section{Conclusion}
A series of methanol-water solutions have been investigated by neutron
diffraction and MD simulation over a range of concentrations, temperatures
and pressures. The diffraction data were analysed using the EPSR technique
which was found to give results qualitatively similar to that from MD simulations,
although there are some differences in detail. A general conclusion of these
studies is that lowering the temperature has the effect of enhancing the
degree of microsegregation between methanol and water that occurs in these
systems. 

More surprisingly, increasing the pressure appears to have the same effect,
which argues against the notion that pressure denaturation of proteins is
caused by water entering the hydrophobic core of a protein: if it did so we
might have expected to see a decrease in the clustering with increased
pressure, not increased clustering. The second shell of water $g_{Ow-Ow}$ appears to expand slightly with increasing
methanol concentration, suggesting a general opening up of the water
structure as the water concentration diminishes: this effect is to be
analysed to identify whether it is primarily a surface effect or proceeds throughout the water. This expansion is if anything reversed on
lowering the temperature or increasing the pressure. We speculate that these
trends could be an indication of the approach to an upper critical solution
boundary, which however is not observed due to the intervening solid phases.

Overall the methanol-water system has proved itself to be a rich source of
phenomena which may be of relevance to situations involving much larger and
more complicated molecules. Methanol and water are ideally suited to the
experimental diffraction and atomistic simulation methodologies due to their
simple molecular forms, and their ready availability in different isotopic
forms. Yet this simple model system can apparently capture much of the
essence of hydrophobicity in aqueous systems in a way that it might occur in
the much larger molecular entities, with mixed hydrophobic and hydrophilic
headgroups of real biological systems. The present results should therefore
help guide the search for possible mechanisms which control molecular
conformation in aqueous solution.

\section*{Acknowledgements}
We acknowledge the technical assistance of John Dreyer, Jonathan Bones and Chris Goodway at the ISIS Facility, without which the experiments described in this paper would not have been possible. Funding from EPSRC is also gratefully acknowledged.

\bibliography{mw_hpltF}
\clearpage

\clearpage

\clearpage

\clearpage

\clearpage

\clearpage

\clearpage

\clearpage

\clearpage

\clearpage

\clearpage

\end{document}